\documentclass{article}
\usepackage{amsmath,
            amsthm,
            hyperref,
            amsfonts,
            cleveref,
            caption,
            graphicx,
            subcaption,
            xcolor,
            natbib}

\title{Spatio-temporal Occupancy Models with INLA}

\author{Jafet Belmont$^{a}$$^{1}$$^{*}$, Sara Martino$^{b}$$^{*}$, Janine Illian$^{a}$,
H{\aa}vard Rue$^{c}$ 
\\
        \small $^{a}$School of Mathematics and Statistics, University of Glasgow, Glasgow, U.K.  \\
        \small $^{b}$Department of Mathematical Scienqces, NTNU, Trondheim, Norway.\\
        \small $^{c}$King Abdullah University of Science and Technology, Thuwal, SA \\\\
        \small $^{1}$Corresponding author: Jafet Belmont; \tt{jafet.belmontosuna@glasgow.ac.uk} \\
         \small $^{*}$Equally contributing authors \\
}

\usepackage{bm}

\begin{document}
\maketitle

\abstract{

 Modern methods for quantifying, predicting, and mapping species distribution have played a crucial part in biodiversity conservation. Occupancy models have become a popular choice for analyzing species occurrence data due to their ability to separate out the observational error induced by imperfect detection of a species, and the sources of bias affecting the occupancy process. However, the spatial and temporal variation in occupancy that is not accounted for by environmental covariates is often ignored or modelled through simple spatial structures as the computational costs of fitting explicit spatio-temporal models is too high. In this work, we demonstrate how INLA  may be used to fit complex spatio-temporal occupancy models and how the \texttt{R-INLA} package can provide a user-friendly interface to make such complex models available to users.

 We show how occupancy models, provided some simplification on the detection process is assumed,  can be framed as latent Gaussian models and, as such, benefit from the powerful INLA machinery.  A large selection of complex modelling features, and random effect models, including spatio-temporal models, have already been implemented in \texttt{R-INLA}. These also become available for occupancy models, providing the user with an efficient, reliable and flexible toolbox.

 We illustrate how INLA provides a flexible and computationally efficient framework for developing and fitting complex occupancy models using two complex case studies. Through these, we show how different spatio-temporal models that include spatial-varying trends, smooth terms, and spatio-temporal random effects can be fitted to aggregated detection/non-detection data. At the cost of  limiting the complexity of the detection model structure, INLA can incorporate a range of rather complex structures in the ecological process of interest and hence, extend the functionality of occupancy models.

 The limitations of occupancy models in terms of scalability for large spatio-temporal data sets remains a challenge and an active area of research. INLA-based occupancy models provide an alternative inferential and computational framework to fit complex spatiotemporal occupancy models. The need for new and more flexible computationally efficient approaches to fit such models makes INLA an attractive option for addressing complex ecological problems, and a promising area of research. 
}


\section{Introduction}

The growing concern of the impact of anthropogenic actions on the environment such as habitat loss, climate change, and land use modification, and of how the distribution of species might be altered  as a result of these  has spurred the development of a wide range of methods for modelling and predicting species presence and abundance in space (and time)  \cite{elith2009species}. 
The resulting species distribution models may be used to estimate population sizes for conservation purposes or to understand habitat preferences and predict species presence or abundance in areas that have not been surveyed or into the future. 

Species distribution models are based on data collected on the presence (or abundance) of species in space and time. Depending on the species and system of interest, available resources and other practical issues such as accessibility, data collection methodology varies among surveys.  Different data collection methods result in different data structures and in turn, require different modelling approaches. However, one thing that most of these surveys have in common is the problem of imperfect detection,  resulting from the fact that some  individuals may be missed as they are elusive or otherwise hard to detect, or only temporarily absent. To avoid bias, imperfect detection has to be accounted for, and detection probabilities have to be estimated as part of the modelling process \citep{cressie2009accounting}.

Occupancy data are data that result from a sampling approach that is relatively simple to implement (as opposed to, e.g., mark capture-recapture) but  still provides a means of accounting for imperfect detection. Here, specific sites are visited by observers repeatedly over time and at each instance a species’ presence and absence is recorded. Occupancy models are then used to  model  jointly both the specie presence in space-time and relative to relevant covariates and the detection probabilities, which also might depend on covariates \cite{mackenzie2002estimating,tyre2003improving,cressie2009accounting,dorazio2005estimating,dorazio2006estimating,kery2010correcting,kery2020applied}. 

A wide range of occupancy models have been developed and applied, from simple  single species models \citep{mackenzie2002estimating, tyre2003improving} to complex spatio-temporal models  \citep{rushing2019modeling,diana2022fast}, and have proven to be a flexible tool that ecologists and conservationists use to study species  distributions.  Increasingly complex  modelling approaches and associated model fitting software have been used to fit these to data. Initially, classical maximum likelihood approaches were a popular choice, in particular during the early 2000's \citep{kellner2014accounting}, for example using packages such as PRESENCE  \citep{hines2006presence} or \texttt{unmarked} \citep{unmarked2011,guillera2017modelling}). 
These libraries account for imperfect detection and allow the use of random effects, but lack functionality to account for spatial and temporal autocorrelation beyond the dependence on covariates.  With increasing model complexity, Bayesian modelling approaches have  become more common,  facilitating useful model expansions, such as incorporating spatial or temporal dependence in the description of the processes governing occupancy and detection \citep{devarajan2020multi}.  Accounting for spatial dependence, in particular, is of crucial importance when modelling the spatial distribution of species is of interest since ignoring that measurements taken at nearby locations are dependent can, in fact, produce biased occupancy estimates \citep{wright2019identifying}.  
Since spatial dependence may be the result of complex unobserved ecological processes driving species distribution that covariates cannot explain it is necessary to include model structures that reflect spatio-temporal dynamics to accounts for autocorrelation not explained by covariates, and that reflect ecological processes such as dispersal limitation or site fidelity, ultimately leading to increased model complexity. 

Fitting spatially explicit occupancy models requires powerful computational machinery. Spatially explicit occupancy models can be implemented in widely-used software packages such as NIMBLE \citep{nimble2017} or Stan \citep{carpenter2017stan}, allowing  spatial random effects to be modelled via conditional autoregressive models (CAR models), where the choice of neighborhood structure can  have a large impact on the results. Despite these packages have proven to be a flexible tool for Bayesian analyses, they were not specifically designed for occupancy models, hence users typically experience long running-times when fitting such models \citep{clark2023gibbs}.  Moreover, while these programming languages are very flexible, they are often not optimized for efficient computation involving dense covariance matrices and also may require relatively advanced programming skills.
On the other hand, bespoke packages such as \texttt{stocc}  \cite{stocc}, \texttt{ubms} \citep{ubms}, \texttt{hSDM} \citep{hSDM} and  \texttt{Rcppocc} \citep{clark2019efficient} provide a user-friendly interface for fitting spatial occupancy models and can implement areal spatial models using (intrinsic) CAR  models. However, these packages cannot fit multivariate  models and may not be adequate for large datasets (e.g.\ data collected at several thousand locations).  Recent attempts to make spatial occupancy models more efficient have focused on using Gaussian Process approximations, for example,  the package \texttt{spOccupancy} \citep{doser2022spoccupancy}  uses Nearest Neighbor Gaussian Processes (NNGP) \citep{diana2022fast} to allow the user to fit models to large spatial datasets.

These bespoke implementations use MCMC algorithms to generate posterior samples, but are typically limited to the supported model types each package provides. As a result, users cannot customize their model structure unless they code their own MCMC  algorithm \citep{ponisio2020one}. 
In addition, MCMC algorithms can be particularly  inefficient for spatial and spatio-temporal  \citep{blangiardo2015spatial,jona2013discussing}. Moreover, large binary occurrence  data in occupancy models can make convergence and chain mixing difficult, even for non-explicit spatial models \citep{northrup2018comment}. 

An alternative approach to MCMC-based algorithms is Integrated Nested Laplace Approximation (INLA) introduced in \cite{rue2009approximate}. Unlike MCMC, which can virtually be applied to any Bayesian model, INLA is a more specialized tool and only applies to the, albeit very wide, class of Latent Gaussian models (LGMs). A LGM consists of three elements: a likelihood model, a latent Gaussian field  and a vector of non-Gaussian hyperparameters. The data are considered conditionally independent given the latent field, and hence a univariate likelihood model describes the marginal distribution of the observations. As in the generalized linear model framework, the mean (or another measure of central tendency) of the observations is linked to a Gaussian linear predictor through a known link function. The core of LGMs lies in its second element: spatio-temporal structures (both discrete and continuous), covariate effects (either linear and non-linear), and other random effects can all be elements of the latent Gaussian field. Generalized linear mixed models (GLMM), generalized additive mixed models (GAMM), spatial and spatio-temporal models  are all examples of LGM \cite{rue2009approximate}.

INLA exploits the Gaussianity of the latent field and uses sparse methods for Gaussian Markov random fields  (GRMF) \citep{rue05_gauss_markov_random_field} that allow for fast computations even for large models. Together with the so called SPDE approach \citep{lindgren2011explicit},  which represents spatially continuous M\'{a}tern Gaussian field as a GMRF, INLA has become a very popular tool that has facilitated
the increased use of Bayesian modelling within the community of applied users (for a recent review see \cite{inlaReview2017, spdeReview2018})

The popularity of the INLA/SPDE approach stems from three sources: the high computational efficiency of the algorithm,  the flexibility that the LGM framework offers, and the fact that the INLA algorithm is implemented in the \texttt{R}-library \texttt{R-INLA}. Thanks to these, INLA has made complex Bayesian modeling more easily available to practitioners. In recent years, many ecological studies have used  the INLA machinery,  and its flexibility and ease of implementation has facilitated the fitting of complex spatio-temporal model \citep{paradinas2015bayesian,Guillen2023}, point processes \citep{Williamson2022,laxton2023balancing}, as well as joint models where different data sources are merged together  \cite{martino2021integration}; this includes integrated species distribution models (IDMs) \citep{simmonds2020more,Paradinas2023,altwegg2019occupancy}.

Occupancy models have, until now, been excluded from the class of models amenable to INLA because of the specific form of the likelihood function, expressed as a function of both occupancy and detection probabilities. 

In this paper we show that, if we are willing to assume a simplified form for one of the two unknown probabilities, the occupancy model can be written in the form of an LGM and,  therefore, benefit from the entire existing INLA machinery. The advantage is that a large class of random effects,  already implemented in \texttt{R-INLA}, also become available for occupancy model. The practitioner no longer needs to either implement bespoke MCMC algorithms for each new study or to be limited to the model choices made by developers of specific occupancy modelling software. 

An added advantage of fitting model within the INLA framework is that measures of model fit such as DIC, WAIC and marginal likelihood \citep{dic2002, waic2013} are readily available. In addition, measures based on leave one out cross-validation (LOOCV) and the recently introduced measures based on leave group out cross-validation (LGOCV) can also be computed without the need to refit the model. The LGOCV is a novel idea introduced in \cite{liu2022leave} and is a better option than LOOCV for  evaluating the predictive performance of models incorporating structured random effects \citep{LGOCV2}.

In this work, we illustrate, through  case studies, how INLA can accommodate a wide variety of already existing occupancy models and extend some of these to more complex scenarios. Using this approach can be useful both as a theoretical framework that can cover many of the specific models and as a practical tool for practitioners that can relate to a single software for their analysis.

\section{Statistical models}

In ecological studies, information on the distribution of a species in space may be gleaned by the presence or absence of individuals of the species across sites. Whilst presence can be confirmed by the detection of individuals, non-detections may be the result of either the species being undetected during the survey or being truly absent at a site. Thus, occupancy models account for non-detection of species even though it is present at a site, by jointly modelling the species' occurrence and detection processes through occupancy and detection probabilities, respectively \citep{mackenzie2002estimating}. In order for these probabilities to be estimated, sites are surveyed at multiple occasions, across which the true occupancy state is assumed to remain constant. For a set of $M$ discrete sites, occupancy data are structured in terms of the detection history of whether a species is detected or not at a given site across multiple visits, i.e.\ $\bm{d}_i= d_{i1},\ldots,d_{iK}$ for the $i$-th site with $K$ visits. Thus, the joint likelihood of occupancy states and detection through time is:

\begin{align}
 L(\bm{\psi},\bm{p}|\bm{d}_1,\ldots,\bm{d}_M) = \prod_{i=1}^{m} \left[\psi_i \prod_j^{K_i} p_{ij}^{d_{ij}} (1-d_{ij})^{1-d_{ij}}\right]\prod_{i=m+1}^M\left[\psi_i\prod_j^{K_i}(1-p_{ij})+(1-\psi_i)\right],
 \label{eq:1}
\end{align}

\noindent
where $\bm{\psi}$ and $\bm{p}$ are the occupancy and detection probabilities, respectively. The first part of Equation (\ref{eq:1}) refers to those $m$ sites where the species is detected at least once throughout the $K$ visits. The second part of Equation (\ref{eq:1}) corresponds to the remaining sites, where the species is not detected at all (i.e.\ $\sum_j y_{ij}  = 0$). If detection probabilities are assumed constant over the $K$ visits (i.e.\ $p_{ij} = p_i \  \forall \  j$), the observational model can then be seen as an aggregated process across the visits. Under this assumption, let $y_i = \sum_j d_{ij}$, i.e.\ the total number of times the species was detected; now the likelihood becomes:

\begin{align}
L(\bm{\psi},\bm{p}|\bm{y}) & = \left[\prod_{i=1}^{m}\psi_i p_i^{y_{i}} (1-p_i)^{K_i- y_{i}}\right]\left[\prod_{i=m+1}^M\psi_i(1-p_i)^{K_i}+(1-\psi_i)\right] \nonumber\\
 & = \prod_{i=1}^{M}\left[\psi_i p_i^{y_{i}} (1-p_i)^{K_i- y_{i}} + I_{[y_{i} = 0]}(1-\psi_i)\right].\label{eq:ZIB}
\end{align}

The likelihood in Equation (\ref{eq:ZIB}) is now factorized into univariate models for the marginal distribution of the observations. This likelihood is a mixture of a binomial density and point mass at $y_i = 0$, i.e.\ there is a probability $\psi_i$ of drawing from a Binomial $(K_i,p_i)$ density, and a probability (1-$\psi_i$) of drawing a zero (i.e.\ a species absence).  If we link \textit{either} the  detection  $p_i$ \textit{or} the occupancy  $\psi_i$ probability to a linear predictor $\eta_i$, using some known link function $g(\cdot)$, while letting the other probability be part of the hyperparameter vector, then the likelihood in Equation (\ref{eq:ZIB})  has a form that is amenable to the INLA framework. If, in addition, we assume a Gaussian distribution for the linear predictor $\eta_i$ then the model can be seen as an LGM. Note that the link function $g(\cdot)$ can be chosen to be a logit, probit or cloglog link.

Placing the occupancy model into the LGM framework gives access to the entire existing INLA machinery that is already implemented in the \texttt{R-INLA} software. 
The model for the parameter that is linked to the linear predictor $\eta_i$ can include spatial effects (both continuous and areal) as well as  smooth effects of covariates and other available random effects (see \texttt{names(inla.models()\$latent)}  for all available random effects models). 
Usually, in the INLA framework, hyperparameters are constant, i.e.\ they do not depend on covariates. That would imply a constant detection probability over the whole study domain (both in the time and space component). As this seem a too restrictive assumption, when implementing the likelihood in Equation (\ref{eq:ZIB}) into \texttt{R-INLA} we have added some extra flexibility allowing for a transformation of the probability $\theta_i = h(p_i)$ (here $h(\cdot)$ can be the logit, probit or cloglog link) to depend linearly on site covariates. No random effect can be included in in this part of the model.
In the context of occupancy models, the main interest typically focuses on the occupancy probability as it associated with the true latent occupancy state. Hence, a most natural choice is to link the linear predictor to the probability of occurrence. The reverse modelling choice is still available and  some practitioners might find this option relevant.
Thus, while maintaining a relatively simple structure within one model sub-component, the other sub-component can flexibly accommodate a wide range of complex terms.

In the next section, we use two case studies to illustrate some of the spatio-temporal models that can be fitted with this approach.  First, we consider two spatial models with independent temporal and spatial (structured and unstructured) random effects. Such models can also be fitted with the  \texttt{spOccupancy} library \cite{doser2022spoccupancy}. However, we  showcase how to include smooth terms in the linear predictor of an occupancy model in addition to the spatial and temporal random effects. A feature that, to our knowledge, is only available in INLA.  Then, we provide a spatio-temporal extension of both types of models where a separable-space time model is fitted, i.e.\ a dynamic spatio-temporal model, where the structured spatial component evolves through time with autoregressive dynamics \citep{cameletti2013spatio}. 
The second case study illustrates how spatially-varying coefficients in occupancy models can be estimated using INLA. Spatially-varying coefficient models (SVCMs) are a flexible class of models  used when the relationship between the response and some observed covariates is not uniform across  space \citep{Gelfand2003,Finley2011}. These models have been by used by \cite{Meehan2019} to quantify the spatial variation in birds populations trends across discretized regions. In this work, we analyse data from the  North American Breeding Bird Survey (BBS) to demonstrate how spatially-varying coefficients  can be estimated in an occupancy model for a set established  birds conservation regions.  Furthermore, we formulate this model in a continuous-space framework and use the proposed  INLA-based  occupancy model to identify significant changes in the continuously and spatially varying trends. 

While the structures of all of the aforementioned models vary widely in complexity, the syntax for fitting such models in INLA varies only slightly. Users, who are already familiar with INLA can easily fit these models or tailor our code to their specific requirement, while new users can benefit from the accessible \texttt{R} code and simulated examples available in the \href{https://ecol-stats.github.io/Occupancy-Models-in-INLA-/website/docs/}{supplementary online material}. This online resource offers a comprehensive explanation of the models under consideration and illustrates how various datasets can be simulated for their subsequent analysis.

\section{Case studies}

\subsection{Modelling the spatio-temporal distribution of  the Red-eyed Vireo in the Hubbard Brook Experimental Forest,  New Hampshire, USA.}

The Hubbard Brook Experimental Forest (HBEF), located in central New Hampshire within the White Mountain National Forest, is a 3600 ha  hydrological research area that has been  used extensively to monitor long-term changes in avian populations \citep{holmes2011avian}.  In this study, we modelled the spatio-temporal distribution of  \textit{Vireo olivaceus} (a.k.a \  Red-eyed Vireo) a migrant species and established breeder in the HBEF. We used  \textit{V. olivaceus} occupancy data obtained from the Valleywide Bird Survey (VBS)  \citep{rodenhouse2019valleywide}. The VBS is a monitoring scheme, that has been  running in HBEF  since 1999, where  transect surveys  have been conducted  each year to collect land bird  point-count  data during the breeding seasons (May-July). A total of 373 locations are  monitored, for each location $i = 1,\dots,373$ and time $t = 2010,\dots,2018$,  the number $K_{it}$ of visits varies between 1 and 3  (see \cite{van2019climate} for more details).

We model  \emph{V. olivaceus} occurrences as:
\begin{align}
z_{it} &\sim \mathrm{Bernoulli}(\psi_{it}) \nonumber \\
& \mathrm{logit}(\psi_{it}) = \eta_{it} \nonumber \\
y_{it}|z_{it} &\sim \mathrm{Binomial}(K_{it},z_{it} \times p_{it})\nonumber\\
\mathrm{logit}(p_{it}) &= \alpha_0 + \alpha_1 \mbox{ mean date of survey}_{it} +  \alpha_2\mbox{mean date of survey}^2_{it},
\label{eq:occ_model1}
\end{align}

where $z_{it}$ indicates the occupancy status at site $i$ at time $t$, while $y_{it}$ indicates how many of the $K_{it}$ visits resulted in a detection.
The occupancy state $z_{it}$ for the $i$-th site in year $t$ is determined by the occupancy probability $\psi_{it}$, defined on the logit scale through the linear predictor $\eta_{it}$. We fitted three separate space-time models for $\eta_{it}$ all described in Table \ref{tab:1}.

\begin{table}[h!]
\centering
\caption{Spatio-temporal structures and covariate effects of the linear predictor fitted in a space-time occupancy model for the  Red-eyed Vireo VBS data.}
\begin{tabular}{|l|l|}
\hline
\multicolumn{1}{|c|}{Model}              & \multicolumn{1}{c|}{Linear predictor $ \eta_{it}$}                                  \\ \hline
Model 1: IID spatial effect                & $\beta_0 + f_1(\mbox{elevation}_i) + f_2(t) + f_3(i)$\\ \hline
Model 2: Continuous spatial model & $\beta_0 + f_1(\mbox{elevation}_i) + f_2(t)  + \omega(i)$\\ \hline
Model 3: Continuous space-time model    & $\beta_0 + f_1(\mbox{elevation}_i) +  \omega(i,t)$\\ \hline
\end{tabular}
\label{tab:1}
\end{table}

All three models have an intercept $\beta_0$  (i.e.\ the logit-scale baseline occupancy probability), and a smooth effect, $f_1(\mbox{elevation})$, of  the standardized elevation at each site. They differ in how the space-time component is treated:  Model (1) includes an autoregressive AR1 effect of time $f_2(t)$ and an unstructured spatial effect, $f_3(i)$, that simply consists of independently and identically distributed (IID) random effects of the locations. Model (2) has the same temporal effect ($f_2(t)$) but explicitly accounts for  spatial correlation through the incorporation of a Mat\'{e}rn Gaussian spatial field, $\omega(i)$. Finally, Model (3) includes a full space-time model $\omega(i,t)$, where changes in time happen according to a first order autoregressive process with spatially correlated innovations (see for example \cite{cameletti2013spatio,spde_book} for details about this model). 
In all three cases, the detection probability $p_{it}$  is assumed to vary according to the date of each survey averaged over  each year. 
All three models can easily be implemented in the \texttt{R-INLA} library without the need for ad hoc functions. Further modifications of the model, for example the use of a linear  instead of a smooth effect of altitude, can easily be implemented. 

\begin{figure}
    \centering
    \includegraphics[width=0.8\textwidth]{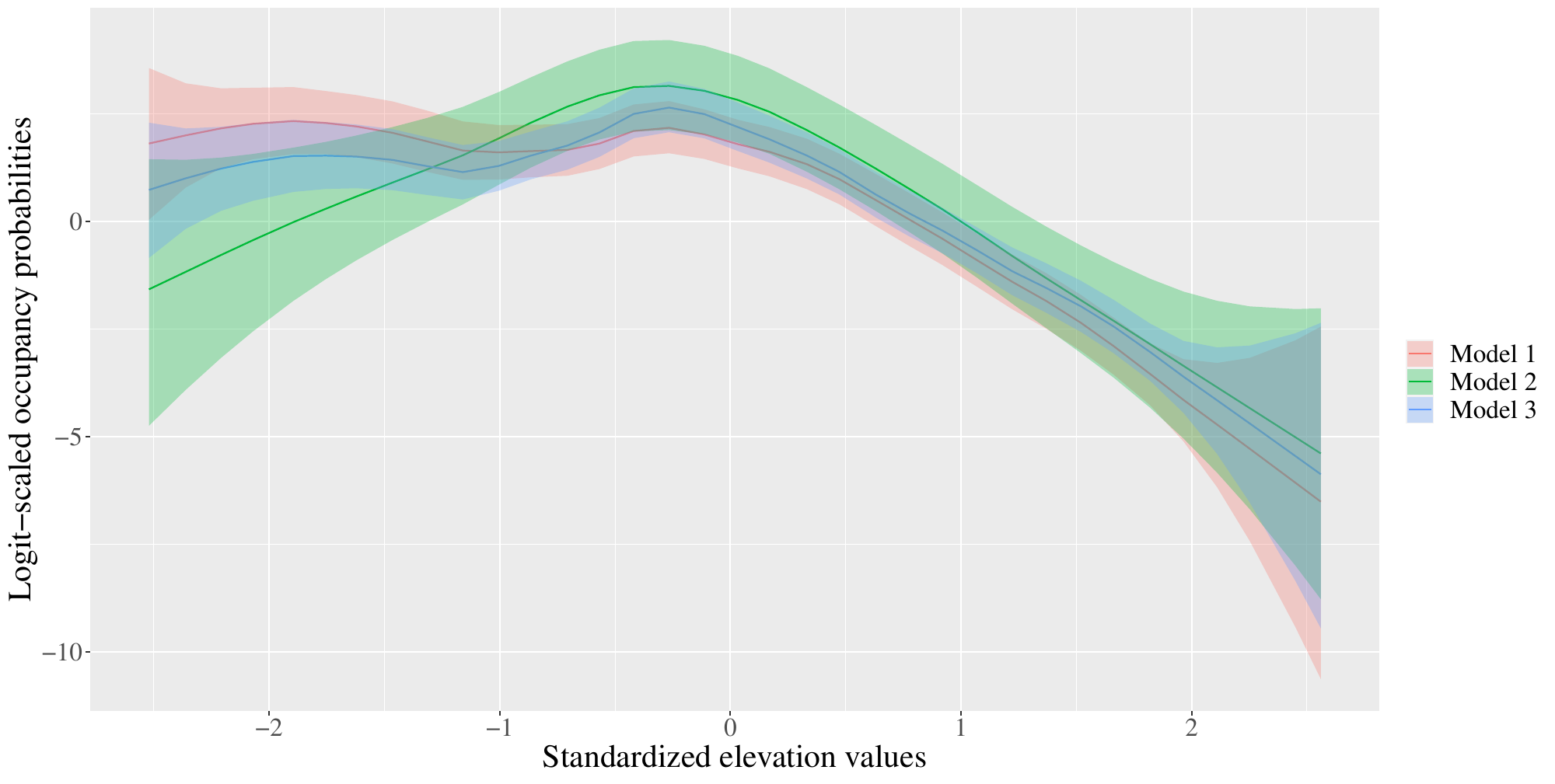}
    \caption{Smooth elevation effect on the logit-scaled occupancy probabilities estimated from the space-time occupancy models for the  Red-eyed Vireo VBS data.}
    \label{fig:1}
\end{figure}

Running times where approximately 2 minutes for the more complex  spatio-temporal Model 3, and less than 10 seconds for the simple \textit{iid} spatial model (Table \ref{tab:2}).
Figure \ref{fig:1} shows the estimated non-linear relationship between (logit) occupancy probability and elevation together with 95\% credibility bands. 
Table \ref{tab:2} reports the posterior means and 95\% credibility intervals for the intercept. All three parameters give similar estimates for this parameter.
The parameters $\alpha_0,\alpha_1,\alpha_2$ of the detection probability were treated as hyperparameters in the implementation. Posterior means and 95\% posterior credible intervals are reported in   Table \ref{tab:2}. Both covariates considered in this part of the model do not appear to be statistically significant.

\begin{table}[h!]
\centering
\begin{tabular}{l|l|l|l}
\textbf{Parameter} & \multicolumn{1}{c|}{\textbf{IID Spatial effect}} & \multicolumn{1}{c|}{\textbf{Continuous spatial model}} & \multicolumn{1}{c}{\textbf{Continuous 
 space-time model}} \\ \hline
 \multicolumn{4}{l}{Occupancy}\\ 
   $\quad \beta_0$& 0.31 ( -0.53, 1.14)& 0.17 (-1.08,  1.36)& 0.46  (-0.21,1.13)\\
 \multicolumn{4}{l}{Detection}\\
$\quad 
 \alpha_0$& \textbf{0.19} (0.13, 0.25)& \textbf{0.18} (0.11, 0.24)& \textbf{0.15} (0.09, 0.22)\\
$\quad 
 \alpha_1$& 0.26 (-0.68,1.07)& 0.26 (-1.25, 1.49)& -0.09 (-0.48, 0.28)\\
$\quad 
 \alpha_2$& -0.53 (-1.34, 0.41)&  -0.52 ( -1.74, 0.99)& -0.16 (-0.53, 0.24)\\
 WAIC& 6694.11& 6679.09&6865.82\\
 DIC&  6714.50& 6688.25&6865.85\\
 Model likelihood&  -3590.46& -3560.36&-3615.87\\
 LGOCV log-score& -1.12& -1.07&-1.09\\
 Run time (sec)& 5.91& 77.80&174.92\\\end{tabular}
\caption{Posterior mean with 95\% Credible Intervals (CrI) for the fixed effects (bold numbers are point-wise estimates for whose CrI do not contain zero) along with the WAIC, DIC, likelihood, log-score and run time for (i) a simple spatial occupancy model with \textit{iid} site-random effects (Model 1); (ii) a spatially explicit occupancy model with continuous spatial processes  (Model 2); (iii) spatio-temporal occupancy model with  spatially continuous process and AR1 component (Model 3).}
\label{tab:2}
\end{table}

\begin{figure}
    \centering
    \includegraphics[width=0.8\textwidth]{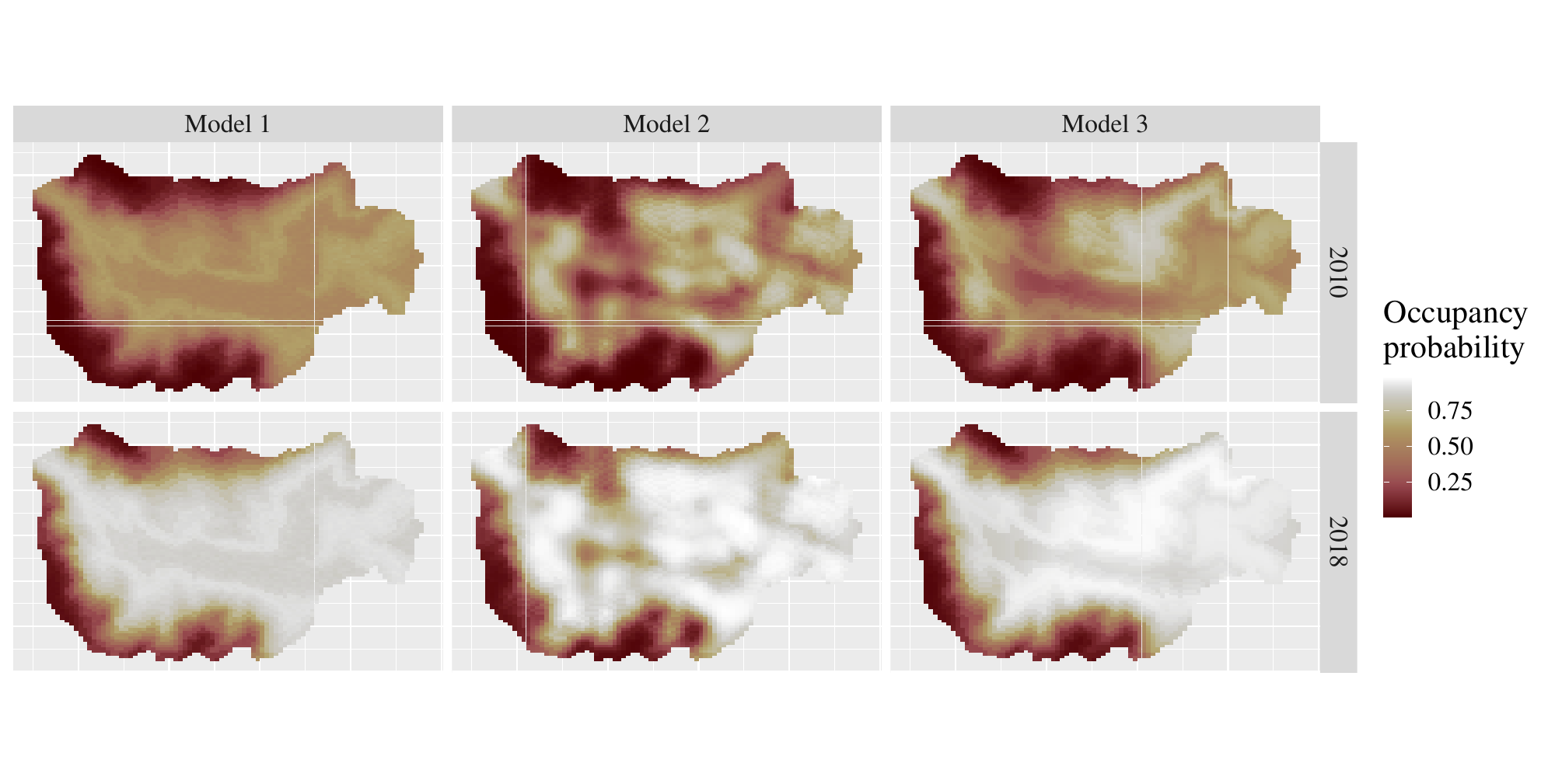}
    \caption{Predicted occupancy probability mean for Red-eyed Vireo across the HBEF  in 2010 and  2018 estimated from three different space-time occupancy models.}
    \label{fig:2}
\end{figure}

\begin{figure}
    \centering
    \includegraphics[width=0.8\textwidth]{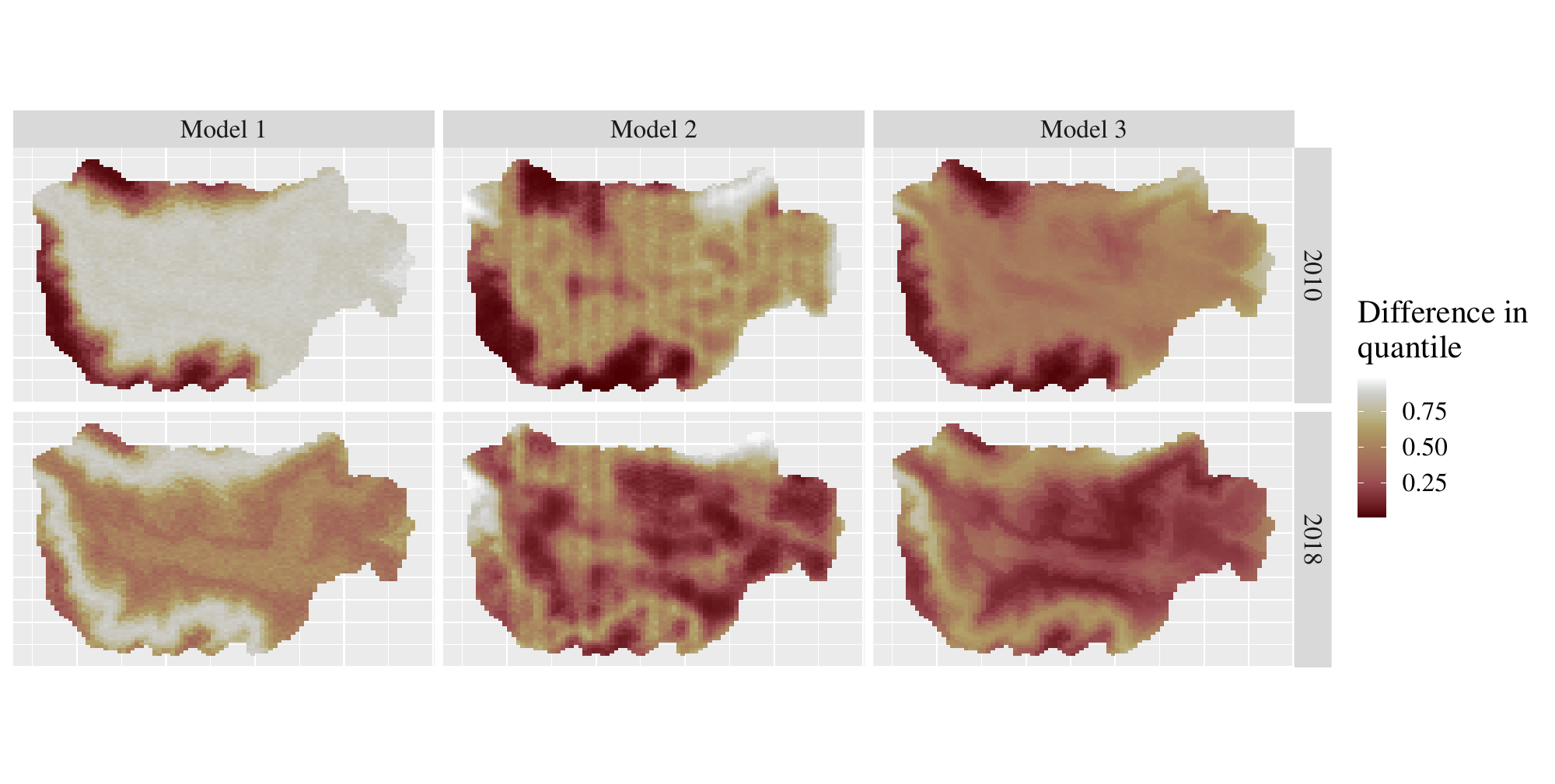}
    \caption{Difference between the 0.975 and 0.025 quantiles of the estimated occupancy probabilities of the Red-eyed Vireo across the HBEF  in 2010 and  2018 estimated from three different space-time occupancy models.}
    \label{fig:3}
\end{figure}

Predictions are also easily available in \texttt{R-INLA} as independent samples from the fitted posterior model can be easily generated \citep{chiuchiolo2023}. Figure \ref{fig:2}  shows the predicted surface of occupancy probabilities  for 2010 and 2018,   for each of the three  models in Table \ref{tab:1}.  The posterior samples can be used to compute different metrics;  for example, Figure \ref{fig:3} illustrates the uncertainty associated with  model predictions by showing the differences between the 0.975 and 0.025 quantiles of the predicted occupancy probabilities for each of the three models. These maps are useful to visualize how prediction uncertainty varies in space: for example it is easy to see that Model (1) is, in general, associated with higher prediction uncertainty than the other two models.

As mentioned, one convenient side effect of the use of INLA is that several model comparison criteria  (e.g., WAIC and DIC as shown in Table \ref{tab:2}) are readily available from the model output. In addition, the recently  developed leave-group-out cross-validation (LGOCV) framework \citep{liu2022leave}, provides cross-validation scores without the need to rerun the model.   We computed the log-score function based on the LGOCV and defined as:

\begin{equation}
    \mathrm{LOGS}_{LGOCV} = \frac{1}{n}\sum_{i=1}^n \pi(Y_i=y_i|\mathbf{y}_{-\mathbb{I}_i}),
\end{equation}
where the  posterior predictive density $\pi(Y_i=y_i|\mathbf{y}_{-\mathbb{I}_i})$ is computed for every observation (testing point) $i$ and an appropriate group $\mathbb{I}_i$ that defines the training set $\mathcal{D}_{-\mathbb{I}_i}$.  In this work, we used INLA's correlation-based automatic construction of the leave out group $\mathbb{I}_i$ (we use the \texttt{inla.group.cv} function and set \texttt{num.level.sets = 3} which controls the degree of independence ), but see the \href{https://ecol-stats.github.io/Occupancy-Models-in-INLA-/website/docs/}{supplementary online material} for an example, where the leave out groups are manually constructed. 
All these measures, for the three considered models,  are reported in Table \ref{tab:2}. All of them disfavour Model 1 while the ranking between Model 2 and 3 is not unique and depends on the chosen criteria.

\subsection{Occupancy model  with spatially-varying coefficients for the Gray Catbird}

The North American Breeding Bird Survey (BBS) is a comprehensive monitoring program devised to track the status and trends of bird populations across North America \citep{BBS2022}. In this work, we use BBS  data on distribution of the  species \emph{Dumetella carolinensis} (a.k.a.\ Gray Catbird) to illustrate how INLA can be used to quantify spatially-explicit trends while correcting for imperfect detection. The data comprise of presence/absence records of the Gray Catbird at $i =1,\dots,1846$ sites sampled at least once between $t = 2000,\dots,2019$. 
The model structure is similar to Equation (\ref{eq:occ_model1}), and again, we showcase  three alternative models for the linear predictor $\eta_{it}$, linked to the probability of occurrence $\psi_{it}$.

\begin{table}[h!]
\centering
\caption{Space-varying time trends in occupancy models for the Gray Catbird BBS Data.}
\begin{tabular}{|l|l|}
\hline
\multicolumn{1}{|c|}{Model}              & \multicolumn{1}{c|}{linear predictor}                                  \\ \hline
Model 1: Fixed time trend& $\eta_{it} = \beta_0 + \omega(i) + \epsilon_t + \beta_t t$\\ \hline
Model 2: Continuously varying time trend & $\eta_{it} = \beta_0 +  \omega(i) + \epsilon_t + \beta_t(i) t$\\ \hline
Model 3: Regional varying time trend& $\eta_{it} = \beta_0 + +\omega(i) + \epsilon_t + \beta_t(\mbox{bcr}_i) t$ \\ \hline
\end{tabular}
\label{tab:4}
\end{table}
The three models are summarised  in Table \ref{tab:4}. Again, all models contain an intercept $\beta_0$, a spatially continuous Gaussian  random field with Mat\'{e}rn covariance structure, $\omega(i)$, and an iid effect of the year $\epsilon_t$. 
They differ in the way a linear time trend is treated. In Model 1, the time trend $\beta\ t$ is constant over the whole space, while in the other two models the time trend varies in space. In Model 2 the time trend varies continuously with time while in Model 3 the time trend varies discretely, by bird conservation region (brc) as defined by the North American Bird Conservation Initiative \citep{BCR2023}).  We use a Mat\'{e}rn continuous Gaussian model for $\beta_t(i)$ in Model 2 and the Besag-York-Molli\'{e} model (BYM) \citep{besag1991bayesian} (an extension to the intrinsic CAR model that contains an  i.i.d.\ model component for the non-spatial heterogeneity) for $\beta_t(\text{bcr}_i)$ in Model 3.

Finally, we define the logit of the  detection probabilities as a linear and quadratic function of the Julian date of the visit, i.e.\ $\mathrm{logit}(p_i) = \alpha_0 + \alpha_1\mbox{day}_i +  \alpha_2\mbox{day}^2_i$.

Table \ref{tab:5} summarizes the posterior means and 95\% credibility intervals for the  baseline occupancy probability ($\beta_0$) and the parameters of the detection model ($\alpha_0, \alpha_1, \alpha_2$) for all three models.  These  estimates are rather similar for the three models. 
The same Table also reports the DIC, marginal likelihood and log-LGOCV score: all three scores marginally prefer the  continuously  varying coefficients. 

\begin{table}[h!]
\centering
\begin{tabular}{l|l|l|l}
\textbf{Parameter} & \multicolumn{1}{c|}{\textbf{Constant spatial trend}} & \multicolumn{1}{c|}{\textbf{Continuously spatial-varying trend}} & \multicolumn{1}{c}{\textbf{Regionally spatial-varying trend}} \\ \hline
 \multicolumn{4}{l}{Occupancy}\\ 
   $\quad \beta_0$& 0.064 (-1.38, 1.51)& 0.30 (-0.84, 1.44)& 0.11 (-1.28,1.50)\\
 $\quad 
 \beta_t$& \textbf{0.17} ( -0.13,-0.21)& ---&---\\
 \multicolumn{4}{l}{Detection}\\
$\quad 
 \alpha_0$& \textbf{0.46} (0.45,0.48)& \textbf{0.46} (0.45,0.48)& \textbf{0.45} (0.46,0.47)\\
$\quad 
 \alpha_1$& \textbf{3.24} (2.95,23.5)& \textbf{3.73} (3.44,3.99)& \textbf{3.26} (2.95,3.58)\\
$\quad 
 \alpha_2$& \textbf{-3.12} (-3.37,-3.12)& \textbf{-3.60} (-3.84,-3.31)& \textbf{-3.13} (-3.45,-2.83)\\
 WAIC& 84062.2& 83848.44 &84018.57 \\
 DIC& 84160.72 & 83978.49 &84121.27 \\
 Marginal  likelihood& -42464.14& -42394.32&-42451.11\\
 LGOCV log-score& -1.591&  -1.589&-1.591\\
 Run time (sec)& 28.53 & 134.11& 43.77\\\end{tabular}
\caption{Posterior mean with 95\% Credible Intervals for the fixed effects (bold numbers are point-wise estimates for whose CrI do not contain zero)  along with the WAIC, DIC, likelihood and run times of the three SVC occupancy models fitted to the Gray Catbird data. }
\label{tab:5}
\end{table}

\begin{figure}
    \centering
    \includegraphics[width=1\linewidth]{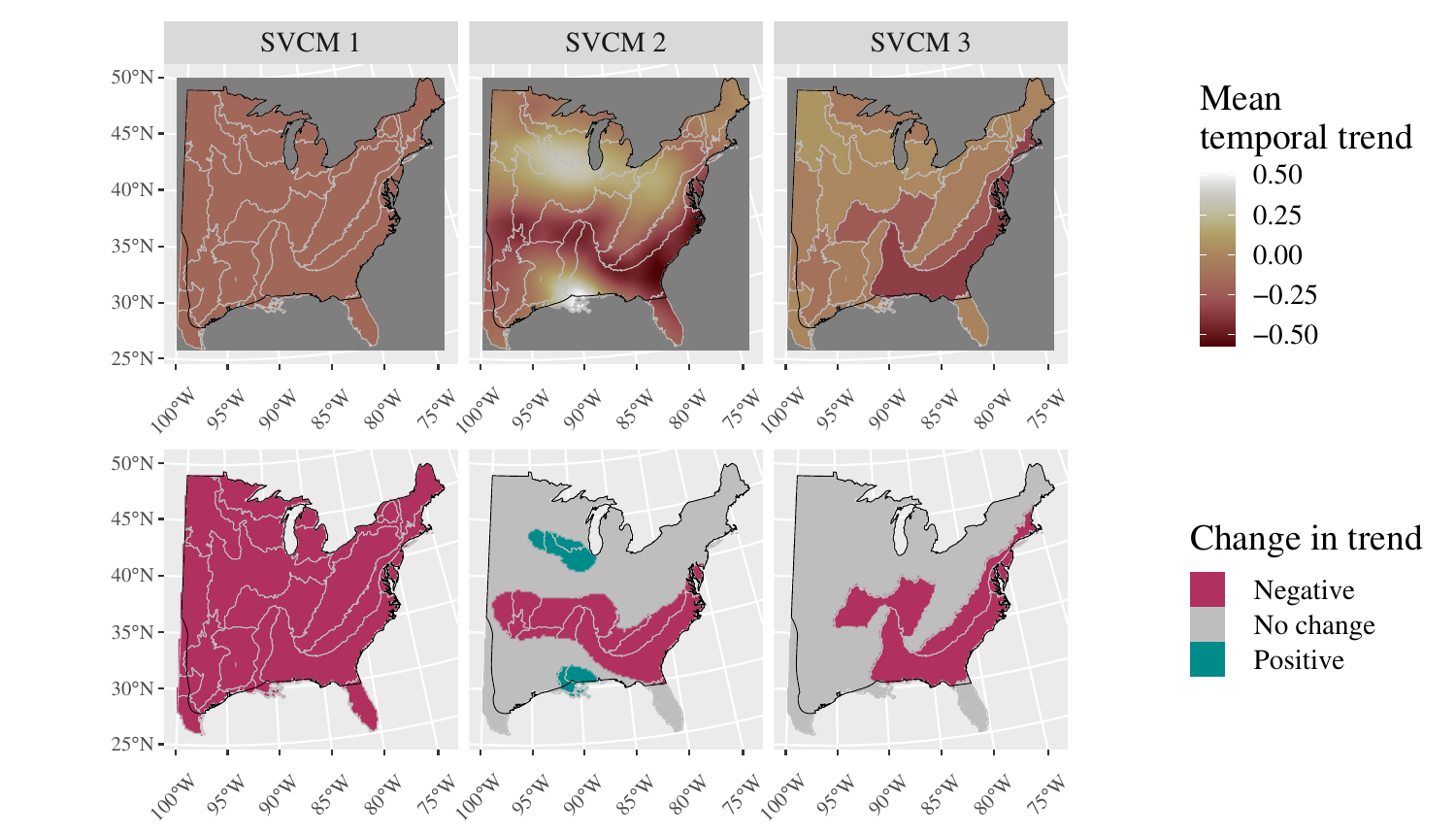}
    \caption{Estimated spatiotemporal trends for the Gray Catbird from 2000-2019.  The first row shows the estimated mean trend for SVCM 1 (fixed constant trend), SVCM 2 (continuously spatial-varying trend), and SVCM 3 (regional spatial-varying trend with BYM structured spatial random effects ). The second row shows significant changes in the trend, i.e. a negative trend if the 95 \% credible interval  < zero, a positive trend if the credible interval was > zero, and no significant change otherwise. }
    \label{fig:4}
\end{figure}

The posterior mean of the estimated temporal trend in space is shown in Figure \ref{fig:4}(upper panel), while areas in space where the credibility interval for the estimated trend does not include 0 are shown in Figure \ref{fig:4} (lower panel). Model 1 estimates a constant and significant negative trend across the whole domain (the estimated constant linear trend for Model 1 is also reported in Table \ref{tab:5}). When we allow the trend to vary in space we get a more nuanced picture: Model 2 identifies ares in the central part of the domain with significantly negative trend but also small areas with significantly positive trend, while Model 3 identifies only areas with a significantly negative trend. Model 2 and Model 3 agree on identifying such areas. 
Visualization of the SVC models predictions over space can be used to inform the monitoring of long-term changes in occupancy trends, which is a major goal in many ecological studies as this allow hot spots relevant for  conservation and managements to be identified \citep{altwegg2019occupancy,outhwaite2020complex,coomber2021using}.

\begin{figure}
    \centering
    \includegraphics[width=1\linewidth]{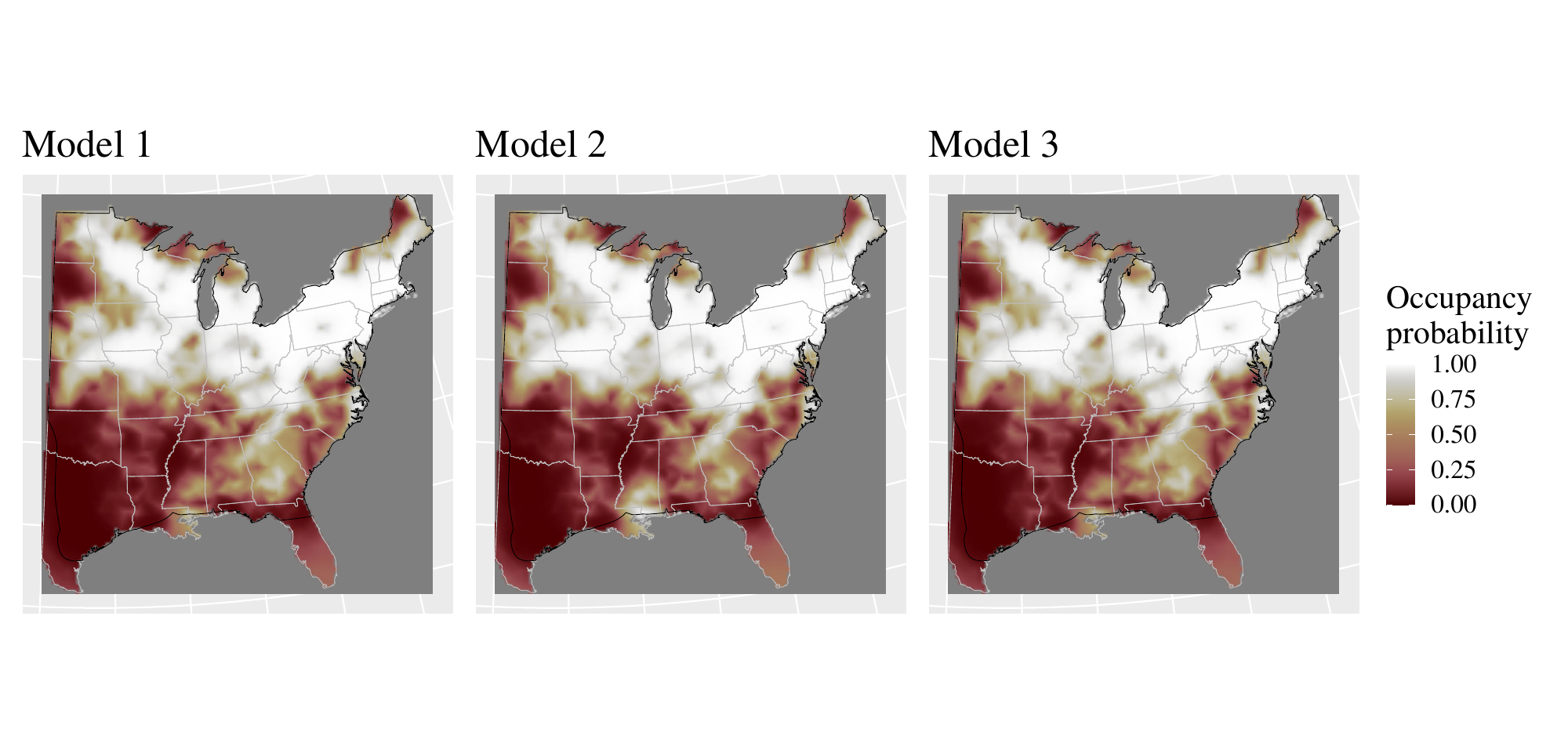}
    \caption{Gray Catbird estimated occupancy probabilities from three SVCMs in 2020.}
    \label{fig:5}
\end{figure}

\begin{figure}
    \centering
    \includegraphics[width=1\linewidth]{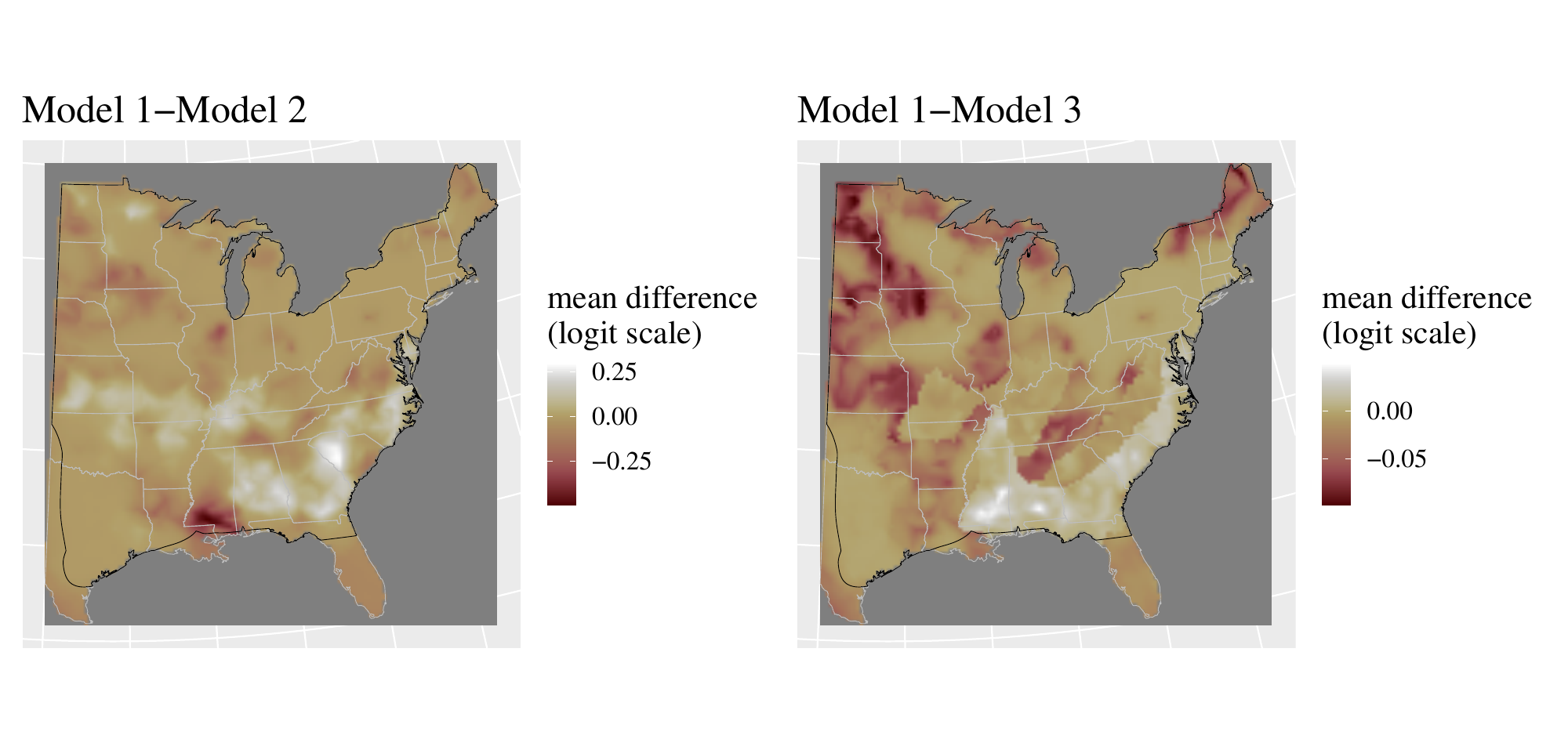}
    \caption{Differences between the estimated logit-scaled mean occupancy probabilities across the SVC occupancy models for the Gray Catbird. On the left, the mean differences between  the fixed trend model and the continuously  spatially varying trend model. On the right, the mean differences between the model with a fixed trend and a regionally varying trend.  }
    \label{fig:6}
\end{figure}

Maps of  posterior means of the predicted occupancy probabilities for Gray Catbird  in 2020 for all three models are reproduced in Figure \ref{fig:5}. The maps appear to be quite similar, however,  spatial differences in logit-scale occupancy probabilities can be identified by comparing the predictions from the fixed trend model (Model 1) against the continuously and regionally spatial-varying trend models (Model 2 and Model 3, respectively; Figure \ref{fig:6} ).

\section{Discussion}

In recent years,  the rise of new technologies and data submission platforms has provided access to many and often multiple data sources (e.g.\ derived from citizen science programs, atlases, museums, and planned surveys)  for species distributions modelling \citep{august2015emerging}. As the volume of recorded ecological data continues to grow,  so does the need for computationally efficient methods for fitting complex spatio-temporal ecological models. Recent developments in spatio-temporal occupancy models have focused on using Gaussian processes (GP) or nearest‐neighbor Gaussian processes (NNGP) approximations while taking the computational advantage of using the P\'{o}lya-Gamma data augmentation scheme \citep{polson2013bayesian}  to improve computational running times. The P\'{o}lya-Gamma data augmentation approach has proven to be a useful method for constructing computationally efficient Gibbs samplers for estimating  occupancy models parameters \citep{clark2019efficient}, and has recently been used for introducing 
spatio-temporal random effects in occupancy models fitted to large data sets   \citep{diana2022fast,doser2022spoccupancy}. However, users interested in  employing this approach must either code their own sampling algorithm or use the predefined set of models provided by a specific package. Additionally, the  P\'{o}lya-Gamma scheme relies on the logit-link function for defining the linear predictor, hence imposing limitations on certain applications (e.g., when aiming to employ the cloglog function to link the linear predictor to the log of the intensity of the underlying Poisson process  \citep{simmonds2020more}). 

INLA is an alternative to MCMC that has already proven to open avenues for groundbreaking progress in many ecological studies \citep[among others]{laxton2023balancing, Paradinas2023, Williamson2022}. In this study, we show that, under some conditions, occupancy models can also be fitted within the INLA framework and therefore benefit from the wide range of models already implemented in the \texttt{R-INLA} software.  The main limitation of the INLA approach is that one cannot use survey-specific covariates in the detection model and that random effects can only be included either in the detection or in the occupancy sub-component. 
In cases where random effects are required in both model components, or  if survey specific covariates need to be accounted for, an MCMC approach might be more appropriate. Nonetheless, \cite{meehan2020}  showed that averaging survey level covariates across sites, yields very similar results for N-Mixture models when comparing INLA with classical or MCMC-based approaches. Our approach is more flexible than  \cite{meehan2020} as it allows: (1) detection probabilities to vary across sites and (2)  random effects to be included in the linear predictor associated with the ecological process of interest.  

Further, it is also important to notice that the log-likelihood for the occupancy models  is not log-concave, therefore, special care has to be taken when formulating the models and choosing the priors. We believe that, for many applied situations, the flexibility and the ease of implementation that the INLA framework offers to the user, will compensate for the slightly more rigid model that one has to adopt for the detection probability.

In recent years,  INLA has become a major computational inferential framework  used by  practitioners and statisticians alike \citep{VANNIEKERK2023107692,van2021new}. Its  popularity is owed not only to the efficient computational machinery, but also to its ongoing development and derivative projects that enhance and expand upon INLA's capabilities  \citep{van2021new}.  Such developments can greatly enhance current occupancy models  theory and applications. For example, prior choice is an important step in Bayesian modelling, penalized complexity (PC) priors  \citep{simpson2017}, as implemented in INLA  has proven to be a useful tool in ecological modelling as a selection tool \citep{sadykova2017bayesian} or as a way of defining the parameters of  the covariance structure of the Mat\'{e}rn field  \citep{laxton2023balancing}. Although PC priors can also be used in contexts outside the INLA framework, the fact that they are implemented in the \texttt{R-INLA} packages, makes them especially convenient for INLA users.  
Furthermore, INLA provides a range of useful tools for model checking and model selection with  model comparison criteria as DIC and WAIC. It also provides the recently introduced LGOCV  strategy, which enables both manual and automatic construction of validation groups while avoiding the need for refitting the model multiple times for each leave-out group to assess the predictive performance of different models in a computationally efficient manner.
In this work, we have presented how INLA can be used to fit a single species occupancy model. However, since the model has been made amenable to the INLA framework, it is easy to build more complex models for multiple or for combining different data sources, where different species or data sets can be represented by different likelihoods that share some of the components in the linear predictor. 

Occupancy models are better suited for protocolized surveys where the sampling design can effectively correct for imperfect detection while minimizing other sources of sampling biases (although occupancy models have also proven to be useful in certain opportunistic surveys \citep{van2013opportunistic}). However, unstructured ecological data without a standardized collection scheme have become more common over recent years, offering a much greater spatio-temporal and taxonomic coverage than protocolized data sets.  Unfortunately, such data sources often consist solely of presence-only records, lacking information on species absence, and are often spatially and temporally  biased towards the choice of the sampling locations \citep{van2013opportunistic}. Thus, integrated distribution models (IDMs) have emerged as a way to combine the information from different data sources, such as those arising  from structured and unstructured surveys, to enhance the accuracy  of the estimates of species distribution models \citep{koshkina2017integrated,simmonds2020more}.   INLA has proven to be powerful platform to fit complex integrated models to multiple data sources with different biases and observational processes \citep{martino2021integration, VILLEJO2023100744}.
Thus, future work may focus on the integration of the occupancy model framework to develop accurate and efficient IDMs. This can also help in cases where site-survey covariates are relevant for the analysis, but not controlled due to the lack of a survey design. For example, data from (i)  planned surveys where site-level covariates can account for observational process biases can be analysed with INLA's occupancy model formulation, and (ii) data from unstructured surveys could be analyzed as a thinned point process as in \cite{yuan2017point, koshkina2017integrated},  while jointly estimating the parameters related to the true species distributions. 

The challenge of scalability for large and complex spatiotemporal datasets is a well-known limitation of occupancy models. Thus,  INLA offers an alternative inferential and computationally efficient framework for handling intricate spatiotemporal occupancy models within an established modelling framework. The appeal INLA has as a modelling framework for occupancy models, lies in its potential to address complex ecological problems by using the machinery already available within the R-INLA software.

\section*{conflict of interest}

We warrant that this manuscript is the original work of the authors listed and has not previously been published. All the authors listed made substantial contributions to the manuscript and qualify for authorship, and no authors have been omitted. We warrant that none of the authors has any conflict of interest in regard to this manuscript

\section*{Supporting Information}

All data sources used for this work are publicly available and have been properly cited in the manuscript. Red-eyed Vireo occurrences data are available from the Valleywide Bird Survey \href{https://doi.org/10.6073/pasta/faca2b2cf2db9d415c39b695cc7fc217}{knb-lter-hbr.178}. Gray Catbrid data  from the  North American BBS can be obtained from \href{https://doi.org/10.5066/P9GS9K64}{https://doi.org/10.5066/P9GS9K64}. R code and simulated examples are available in the \href{https://ecol-stats.github.io/Occupancy-Models-in-INLA-/website/docs/}{online supplementary material}.


\bibliographystyle{plainnat}

\bibliography{sample}

\end{document}